\newcounter{saveeqn}
\newcommand{\alpheqn}{\setcounter{saveeqn}{\value{equation}}
\stepcounter{saveeqn}\setcounter{equation}{0}
\renewcommand{\theequation}
 {\mbox{\arabic{saveeqn}$\,$\alph{equation}}}}

\newcommand{\reseteqn}
{\setcounter{equation}{\value{saveeqn}}
\renewcommand{\theequation}{\arabic{equation}}}

\documentclass[twocolumn,showpacs,preprintnumbers,amsmath,amssymb]{revtex4}

\usepackage{graphicx}
\usepackage{dcolumn}
\usepackage{bm}

\begin{document}
\preprint{1}
\title{Liquid-crystalline Casimir effect in the presence of a patterned substrate}
\author{F. Karimi Pour Haddadan$^*$, F. Schlesener, and S. Dietrich}
\vspace{4.0cm}
\address{\mbox{}\\Max-Planck-Institut f${\ddot u}$r Metallforschung, Heisenbergstr.\,3, D-70569 Stuttgart, 
Germany,\\and\\
Institut f${\ddot u}$r Theoretische und Angewandte Physik, Universit${\ddot a}$t Stuttgart,
Pfaffenwaldring 57, D-70569 Stuttgart, Germany}
\date{\today}   
\draft
\begin{abstract}
We consider a nematic liquid crystal confined by two parallel planar interfaces, one being laterally homogeneous and the other provided by a substrate endowed with a periodic chemical stripe pattern. Based on continuum theory we analyze the influence of the lateral pattern on the liquid-crystalline Casimir force acting on the interfaces of the nematic cell at distance $d$ due to the thermal 
fluctuations of the nematic director. For $d$ much larger than the pattern periodicity, the influence of the patterned substrate can be described by a homogeneous, effective anchoring energy. By tuning this parameter we recover previous results for the liquid-crystalline Casimir force. For the general case, i.e., smaller separations, we present new numerical results.  
\end{abstract}
\maketitle
\section{Introduction}
Liquid crystals in general are sensitive over a wide spatial range to the anchoring
conditions of confining interfaces. This holds also for lateral variations of anchoring conditions
generated either by surface topography~\cite{barbero,bar,brown1,brown2,gre,pedro,sheng} or patterning
~\cite{kamei,sheng6,sheng7,rosenblatt,kitson,kim,kondrat1,ishino,hans,schadt,kondrat2}
giving rise to numerous possible applications. In recent years, the influence of the structured 
substrates on the properties of liquid crystals has been studied and it has been demonstrated that 
to a certain extent such nontrivial geometries may optimize the performance of electro-optical 
liquid-crystalline devices. For instance, a four-domain twisted nematic liquid crystal display
provides a wide viewing angle with no gray scale inversion\cite{perf} and using multistable nematic
liquid-crystal devices with micropatterned substrate alignments reduces the energy
consumption~\cite{brown2,pedro,ishino,kim}.

The influence of the anchoring on the liquid crystal order parameter translates into an effective
interaction between the confining interfaces which may be provided either by true solid substrates
or by an adjacent vapor phase where the former case can support permanent lateral structures. 
Here we consider liquid crystals deep in their nematic phase where the orientational order is 
described by a director field with long-ranged correlations. In case of any mismatch between 
the prescribed alignment of the bulk and the substrates, the director structure may not be uniform.
In such a case, the free energy of the system typically exhibits several, metastable, minima 
which upon a change in the parameters of the system\,\,--\,\,such as 
the film thickness, external, or internal forces\,\,--\,\,may turn into the global minimum 
resulting in a structural phase transition~\cite{pikin}. This makes the stability of the 
equilibrium configuration geometry dependent. In this context one should keep in mind that 
perturbative approaches may miss the occurrence of first-order structural phase transition. We shall 
consider the case of frustrated systems in which, however, the director structure remains uniform up
to a critical thickness $d_c$. Within such a regime the liquid-crystalline mediated effective 
interaction between the confining interfaces due to the thermal fluctuations of the director 
adds to background contributions due to structural forces arising from presmectic 
layering\,\cite{horn,degennes,ziherlp} and enhanced ordering near the substrates\,\cite{horn}
and due to dispersion forces\,\cite{isra} which exhibit only a weak temperature dependence.

Using the continuum Frank free energy, we study the fluctuation-induced 
interaction\,\,--\,\,the so-called liquid-crystalline Casimir effect\,\,--\,\,between two parallel
interfaces where one is periodically patterned and the other one is homogeneous.
We consider a periodicity in the local anchoring energy and model the liquid crystal-substrate 
interaction by the Rapini-Papoular surface free energy. We investigate the modification of the 
fluctuation-induced force compared with its behavior for substrates with uniform anchoring 
conditions~\cite{ajdari} as a function of the pattern periodicity $\zeta$ and the characteristic
length of the pattern $\zeta_a$ (see Fig.\,\ref{geo55.eps}). 

Two model systems are considered. One consists of a substrate with a pattern characterized by 
homeotropic anchoring of alternating strengths facing a second substrate at a distance $d$ which is
characterized by a uniformly strong homeotropic anchoring. For this system the mean director is 
constant at any separation. By changing the boundary condition via the change of the patterning
ratio $\zeta_a/\zeta$, the character of the force changes. Depending on whether the boundaries
are effectively similar-nonsimilar or similar-similar the force is repulsive or 
attractive, respectively. For certain values of $\zeta_a/\zeta$ and of the reduced 
distance $d/\zeta$, the liquid-crystalline Casimir force vanishes. In the second system
the pattern consists of alternating stripes of homeotropic and degenerate planar anchoring while
the upper substrate still exhibits strong homeotropic anchoring. In this case there is the 
possibility of texture formation~\cite{hans}. However, for separations smaller than a critical
one the director structure is indeed uniform~\cite{lud}. For those ranges of the model parameters
for which the director is constant (see, c.f., Subsec.\,\ref{III}), we find a nonmonotonic 
behavior for the fluctuation-induced force.  

There are several techniques that can be used to create periodic anchoring conditions
such as photo-alignment~\cite{ishino}, the selectively thiol-functionalized 
photo-orientation~\cite{hans}, and atomic force lithography~\cite{rosenblatt}.
In the latter case patterning at the nano-meter scale is reached which allows one to investigate
more efficiently the influence of the patterned substrate on the liquid crystal.
The results of our study, for example, might be helpful in designing thin patterned liquid 
crystalline films for which the fluctuation-induced force plays a role for the stability of the film.

In Sec.\,\ref{I} we describe the system and the formalism that we apply for calculating the 
fluctuation induced effective interaction. In Subsec.\,\ref{II} we investigate the force in the 
presence of a pattern of alternating anchoring strengths and in Subsec.\,\ref{III} with a pattern
of competing anchoring conditions. We present analytical results for patterns of small and 
large scales and numerical results for patterns at intermediate scales. In Sec.\,\ref{IV} we 
summarize our results.
\section{System and theoretical model}
\label{I}
\begin{center}
\begin{figure}
\includegraphics[height=.92\linewidth,angle=0]
{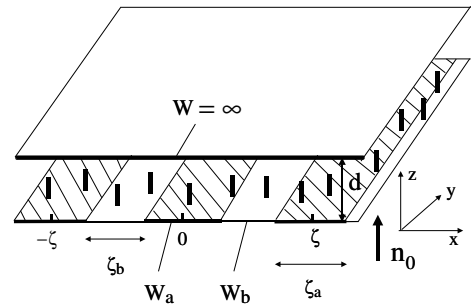}
\vspace{-2.5cm}
\caption{The geometry of the nematic cell with cross section $A$ and volume $V=Ad$. The upper 
boundary is characterized by strong homeotropic anchoring. The lower substrate is patterned. The
pattern consists of periodic stripes of anchoring energies per area $W_a$ and $W_b$ with widths
$\zeta_a$ and $\zeta_b$, respectively, so that $\zeta=\zeta_a+\zeta_b$. The easy 
directions (see Eq.\,(\ref{eq6})) at both boundaries are normal to the interfaces and we 
consider such values of $W_a$ and $W_b$ for which the thermal average of the director 
field ${\bf n}_0$ is homogeneous.}
\label{geo55.eps}
\end{figure}
\end{center}

We consider a liquid crystal in a nematic phase and confined by two parallel planar interfaces at
a distance $d$ (see Fig.~\ref{geo55.eps}). The liquid crystal is described by the Frank free 
energy~\cite{deg}
\begin{eqnarray}
F=\frac{1}{2}\int_{V}{\rm d}^3x\Big[K_1(\mbox{\boldmath $\nabla$}\cdot{\bf n})^2
+K_2({\bf n}\cdot
\mbox{\boldmath $\nabla$}\times{\bf n})^2
\nonumber\\+K_3({\bf n}\times
\mbox{\boldmath $\nabla$}\times{\bf n})^2\Big],
\label{eq0}
\end{eqnarray}
where {\bf n} denotes the director of the liquid crystal, $V$ is the nematic volume, and
$K_1$, $K_2$, and $K_3$ are the splay, the twist, and the bend elastic constants, respectively.
The interaction between the liquid crystal and the substrate is modeled by the Rapini-Papoular
surface free energy~\cite{rapini} given by
\begin{equation}
F_{s}=-{1\over 2}\int_{A} {\rm d}^2x \,W({\bf x})\,({\bf n}\cdot{\bf e})^2,
\label{eq6}
\end{equation}
where ${\bf x}=(x,y)$ denotes the lateral coordinates of Cartesian 
coordinates ${\bf r}=({\bf x},z)$, A is the surface area, $W$ is the anchoring energy per 
area, and ${\bf e}$ is the easy direction, i.e., the preferred direction of the director at 
the substrate if $W>0$. For $W<0$ the director prefers the direction perpendicular to ${\bf e}$. 
In the following we restrict the discussion to the case that ${\bf e}$ is perpendicular to both
substrates, which leads to homeotropic anchoring for $W>0$ and to degenerate (i.e., with no 
preferred azimuthal angle) planar anchoring for $W<0$.
On the lower substrate located at $z=0$, we assume that the anchoring energy varies periodically
along the $x$-direction. The pattern consists of alternating stripes of anchoring energies per 
area $W_a$ and $W_b$. The substrate remains translationally invariant in the y-direction.
On the upper substrate located at $z=d$ we assume uniformly strong homeotropic anchoring
(see after, c.f., Eq.\,(\ref{bound})). 

In general, the local director field is given by
\begin{equation} 
{\bf n}({\bf x},z)={\bf n}_0({\bf x},z)+\mbox {\boldmath $\delta$}({\bf x},z)
\label{eq1p}
\end{equation}
where ${\bf n}_0({\bf x},z)$ is the thermal average of the director and
$\mbox {\boldmath $\delta$}({\bf x},z)$ is the fluctuating part with vanishing thermal average
$\langle \mbox {\boldmath $\delta$}\rangle=0$.
\subsection{Mean-field behavior}
First we discuss the mean-field solution ${\bf n}_0({\bf x},z)$ of the director field. In the case
of homeotropic anchoring everywhere on the lower substrate, the uniform solution ${\bf n}_0=(0,0,1)$
is the equilibrium configuration. On the other hand in case of planar anchoring everywhere on the 
lower substrate, the liquid crystal is subject to competing surface interactions at the top and 
the bottom. For this so-called hybrid cell\,\,--\,\,a cell with uniform homeotropic and uniform
planar anchoring on each substrate\,\,--\,\,it has been shown that the substrate whose anchoring 
is stronger can impose a uniform director configuration up to a critical separation between the
plates~\cite{barberi}. In the case of periodic pattern of homeotropic and planar anchoring on 
the substrate (Subsec.\,\ref{III}), the full phase diagram of the system within mean field theory
and the structural phase transition between a uniform director configuration and a distorted one 
can be studied by means of numerical minimization of the free energy functional~\cite{kondrat1,lud}.
However, one can naively expect that in this system the tendency to form a uniform director field
is enhanced with respect to the hybrid cell due to the presence of the interlaced homeotropic stripes.
In the following we restrict the discussion to separations smaller than the critical separation 
$d_c$ for which the director configuration is uniform and focus on the fluctuations.
\subsection{Fluctuations of the director}    
Next we consider fluctuations around the uniform director ${\bf n}_0=(0,0,1)$. 
Since the director ${\bf n}({\bf x},z)$ is a unit vector the fluctuations can be described by
$\mbox{\boldmath $\delta$}=(\delta_x,\delta_y,-1+\sqrt{1-\delta_x^2-\delta_y^2})
\simeq (\delta_x,\delta_y,-\delta_x^2/2-\delta_y^2/2)$,
where $\delta_x({\bf x},z)$ and $\delta_y({\bf x},z)$ are the two independent components.
According to Eq.\,(\ref{eq0}) and within the one-constant approximation, the bulk contribution to 
the statistical weight $\exp\,\left(-\beta\,{\cal H}_{\rm bulk}[{\bf n}]\right)$ for a director
configuration ${\bf n}$ is given by ${\cal H}_{\rm bulk}[{\bf n}]={\cal H}_{\rm bulk}[\delta_x]
+{\cal H}_{\rm bulk}[\delta_y]$ with
\begin{equation}
{\cal H}_{\rm bulk}[\nu]={K\over 2}\int_{V} {\rm d}^3x\left[\nabla \nu({\bf x},z)\right]^{2},
\label{eq1}
\end{equation}
where $\beta^{-1}=k_BT$ is the thermal energy and $K$ is the effective elastic 
constant~\cite{aside1}. Eq.\,(\ref{eq1}) amounts to considering Gaussian fluctuations, i.e., 
the Hamiltonian is quadratic in $\nu({\bf x},z)$. This is expected to give a qualitatively correct
description of the system except near an incipient structural phase transition. 

As a local contribution the surface interaction is evaluated at the interfaces $z=0$, $z=d$.
The lower substrate is characterized by the patterning function 
\begin{equation}
a(x)=\sum_{k=-\infty}^{\infty}\Theta \Big(x-k\zeta+{\zeta_a\over 2}\Big)\Theta 
\Big(k\zeta+{\zeta_a\over 2}-x\Big),
\label{eqa}
\end{equation}
where $\Theta(x)$ is the Heaviside step function, $\zeta$ is the periodicity, $\zeta_a$ is the 
width of the stripe characterized by $W_a$, and $\zeta_b=\zeta-\zeta_a$ is the width of the 
stripe characterized by $W_b$ (Fig.\,\ref{geo55.eps}). The stripes form sharp chemical steps between
them. The function $a(x)$ is one at the regions characterized by $W_a$ and zero elsewhere.
Accordingly, for this model the surface interaction [Eq.\,(\ref{eq6})] disregarding constant terms
is given by 
${\cal H}_{\rm surf}[{\bf n}]={\cal H}_{\rm surf}[\delta_x]+{\cal H}_{\rm surf}[\delta_y]$
with
\begin{eqnarray}
{\cal H}_{\rm surf}^{z=0}[\nu]={1\over 2}\Big[W_a\int_{A}{\rm d}^2x\, \big(\nu ({\bf x},z=0)\big)^2
a(x) \nonumber\\
+W_b\int_{A}{\rm d}^2x\, \big(\nu ({\bf x},z=0)\big)^2 (1-a(x))\Big]\
\label{eq3}
\end{eqnarray}
for the lower substrate. We assume homogeneous anchoring on the upper boundary so that
\begin{equation}
{\cal H}_{\rm surf}^{z=d}[\nu]={1\over 2}W\int_{A} {\rm d}^2x\,\big(\nu({\bf x},z=d)\big)^2.
\label{eq9}
\end{equation}
Minimization of the Hamiltonian 
${\cal H}[\nu]={\cal H}_{\rm bulk}[\nu]+{\cal H}_{\rm surf}^{z=0}[\nu]+{\cal H}_{\rm surf}^{z=d}
[\nu]$ leads to the
following boundary conditions:
\alpheqn
\begin{eqnarray}
-K\partial_z \!\!\!\!&\nu&\!\!\!\!({\bf x},z)+W_a \nu({\bf x},z)\,a(x)\nonumber\\
&+&W_b \nu({\bf x},z)\,(1-a(x))=0, \; z=0,\label{eq:bca}
\end{eqnarray}
\begin{equation}
K\partial_z \nu({\bf x},z)+W\nu({\bf x},z)=0,\; z=d,\label{eq:bcb}
\end{equation}
\reseteqn
\!\!\!\!where $\nu$ is either $\delta_x$ or $\delta_y$. After integration by parts in 
Eq.\,(\ref{eq1}) and using the boundary conditions given by Eqs. (\ref{eq:bca}) and
(\ref{eq:bcb}), the Hamiltonian ${\cal H}[\nu]$ reduces to 
${\cal H}[\nu]=-{K\over 2}\int_{V}{\rm d}^3x\,\nu({\bf x},z)\triangledown ^2\nu({\bf x},z)$.
In terms of the so-called extrapolation lengths $\lambda_a$ and $\lambda_b$
\begin{equation}
\lambda_{a(b)}={K\over W_{a(b)}},
\label{bound}
\end{equation}
the boundary condition [Eq.\,(\ref{eq:bca})] on the patterned substrate $z=0$ reads 
${\cal A}_{1}({\bf x},z)=-\lambda_b \partial_z \nu({\bf x},z)+
\left(1+{\lambda_b-\lambda_a\over \lambda_a}a(x)\right)\nu({\bf x},z)=0$.
Assuming strong homeotropic anchoring $W=\infty$ $(\lambda=0)$ at the upper boundary, 
Eq.\,(\ref{eq:bcb}) leads to the Dirichlet boundary condition 
${\cal A}_{2}({\bf x},z=d)\equiv \nu({\bf x},z=d)=0$.

As an aside, we note the relation between the present model and those for surface critical phenomena. 
Rescaling the fluctuating field by $K/k_BT$, 
the Hamiltonian in Eq.\,(\ref{eq1}) can be written as
${\cal H}_b[\varphi]=k_BT\int_{V}{\rm d}^3x\,{1 \over 2}(\nabla \varphi)^2$ and the surface
interaction in Eq.\,(\ref{eq9}) is represented by ${\cal H}_S[\varphi]=
k_BT\int_{A}{\rm d}^2x\,{c \over 2}\varphi^2$, where $c=W/K$ is the inverse extrapolation length
of the critical order parameter profile at a surface~\cite{crit}.
The limiting cases $c=\infty$ and $c=0$ correspond to Dirichlet and v. Neumann boundary
conditions, respectively. The bulk Hamiltonian for a system close to the critical point also 
includes the terms ${\tau\over 2}{\varphi}^2$ and ${u\over 24}{\varphi}^4$ of which the former one
vanishes at the critical point. In this sense the present study corresponds to discussing, within 
the Gaussian approximation, a slab of a system at bulk criticality confined by planar surfaces one
of them endowed with a pattern of the extrapolation length.
\subsection{The fluctuation-induced force}
We employ the path integral method introduced by Li and Kardar for calculating the partition
function of the system~\cite{li91,golestanian} which amounts to integrate over all configurations
of the fluctuating field weighted by the Boltzmann factor and subject to the boundary conditions.
We impose the boundary conditions by inserting delta functions into the path integral.
Thus the partition function $Z$ of the field $\nu$ reads
\begin{eqnarray}
Z=\!\!\!&\int& \!\!\!\!\!\!{\cal D}\nu ({\bf r}) ~ e^{-{\cal H}[\nu]/(k_BT)}\nonumber\\
\times\!\!&\mbox{\Large $\Pi$}_{\bf x}&\!\!
\mbox{\large $\delta$} \Bigl({\cal A}_{1}
({\bf x},z=0)
\Bigr)\mbox{\large $\Pi$}_{\bf x}
\mbox{\large $\delta$}\Bigl(
{\cal A}_{2}({\bf x},z=d)\Bigr)\
\label{Z}
\end{eqnarray}
with the functional integral defined via a discretization on a lattice $\{{\bf r}_n\}$ in the limit
of a vanishing lattice constant:
$\int {\cal D}\nu ({\bf r})\equiv 
\prod_{n}\int_{-\infty}^{\infty}{{\rm d}\nu ({\bf r}_n)\over \sqrt{2\pi}}$\,\cite{lub}. Using the 
integral representation of the delta function,
\begin{equation}
\mbox{\Large $\Pi$}_{\bf x}\mbox{\large $\delta$} \Big({\cal A}_{\alpha}({\bf x})\Big)=
\int {\cal D}\Psi_{\alpha} \,\exp\left(i \int 
{\rm d}^{2}x\, \Psi_{\alpha}\,{\cal A}_{\alpha}\right), \, \alpha=1,2,
\label{delta}
\end{equation}
and performing the Gaussian integral over the field $\nu$, we obtain
\begin{equation}
Z={\cal N}\,\int \prod_{\alpha=1}^{2}{\cal D}\Psi_{\alpha}~e^{-{\cal H}_{\rm eff}}\;,
\label{path}
\end{equation}
where $\Psi_{\alpha=1,2}$ are auxiliary fields defined at $z=0$ and $z=d$, respectively, ${\cal N}$ 
is a factor independent of $d$, and the effective interaction reads
\begin{equation}
{\cal H}_{\rm eff}={\sum_{\alpha,\beta=1}^{2}\int
{\rm d}^2 x\int {\rm d}^2x{'}\Psi_{\alpha}({\bf x}) 
M_{\alpha,\beta}({\bf x},{\bf x'}) \Psi_{\beta}({\bf x}{'})},
\label{heff}
\end{equation}
where $M$ is regarded as a matrix both with respect to the indices $\alpha$, $\beta$ and the 
coordinates ${\bf x}$, ${\bf x}{'}$: 
\begin{widetext}
\begin{eqnarray}
&M_{11}({\bf x},{\bf x}{'})=\Big\{[1+{\lambda_b-\lambda_a\over \lambda_a}a(x)]
[1+{\lambda_b-\lambda_a\over \lambda_a}a(x')]
+{\lambda_b (\lambda_b-\lambda_a)\over \lambda_a}
[a(x)-a(x')]\partial_z-\lambda^2_b\partial^2_z\Big\}
G({\bf x}-{\bf x'},z-z{'})\Big |_{z=z{'}=0}\nonumber\\
&M_{12}({\bf x},{\bf x}{'})=\Big(1+{{\lambda_b-\lambda_a}\over \lambda_a}a(x') 
-\lambda_b\partial_{z'}\Big)
G({\bf x}-{\bf x'},z-z{'})\Big |_{z=d,z{'}=0}\nonumber\\
&M_{21}({\bf x},{\bf x}{'})=\Big(1+{{\lambda_b-\lambda_a}\over \lambda_a}a(x) -\lambda_b
\partial_{z}\Big) 
G({\bf x}-{\bf x'},z{'}-z)\Big |_{z{'}=d,z=0} \nonumber\\
&M_{22}({\bf x},{\bf x}{'})=G({\bf x}-{\bf x'},z-z{'})\Big |_{z=z{'}=d}\
\label{M}
\end{eqnarray}
\end{widetext}
\!\!\!where $G({\bf r},{\bf r}{'})
={k_BT\over 4\pi K\,|{\bf r}-{\bf r}{'}|}$
is the bulk two-point correlation function in three dimensions defined by
${K\over k_BT}\triangledown^2 G({\bf r}-{\bf r}{'})=-\delta ({\bf r}-{\bf r}{'})$.

Due to the symmetries in the $xy$-plane it is useful to switch to the Fourier transformed quantities.
We note that if the patterning function $a(x)=1$, i.e., for a homogeneous 
substrate ($\lambda_a=\lambda_b)$, the matrix $M$ is diagonal in the lateral Fourier 
space $({\bf p},{\bf q})$. However, here the patterning function is piecewise either one or 
zero [Eq.\,(\ref{eqa})]. Due to the periodicity of the patterning along 
the $x$-direction, i.e., $a(x)=a(x+\zeta)$, and the translational invariance along 
the $y$-direction, the matrix $M$ in the lateral Fourier 
space, $M({\bf p},{\bf q})=\int \int {\rm d}^2x\;{\rm d}^2x' M({\bf x},{\bf x}')
e^{i{\bf p}\cdot {\bf x}}e^{i{\bf q}\cdot {\bf x}'}$, has the following form:
\begin{eqnarray}
M({\bf p},{\bf q})&=&(2\pi)^2\,\mbox{\large $\delta$}(p_y+q_y)\sum_{m=-\infty}^{\infty}N_m(p_x,p_y)
\nonumber\\&\times\mbox{\large $\delta$}&\!\!\!\! \left(p_x+q_x+
{2 \pi m\over \zeta}\right)
\label{eq2}
\end{eqnarray}
with the $(2\times 2)$ matrices $N_m$ given by
\begin{widetext}
\begin{equation}
N_0=
\left (
\begin{array}{cc}
[(1+{\lambda_b-\lambda_a\over \lambda_a}{\zeta_a\over \zeta})^2
-\lambda^2_b\partial^2_z]
G(p,z-z{'})\big |_{z=z{'}=0}+\phi_0&\;~~~(1+{{\lambda_b-\lambda_a}\over \lambda_a}
{\zeta_a\over \zeta}
 -\lambda_b\partial_{z'})
G(p,z-z{'})\big |_{z=d,z{'}=0}\\[3mm]
\,
(1+{{\lambda_b-\lambda_a}\over \lambda_a}{\zeta_a\over \zeta}
-\lambda_b\partial_{z}
)G(p,z{'}-z)\big |_{z{'}=d,z=0}&\;~~~G(p,z-z{'})\big |_{z=z{'}=d}
\end{array}
\right )
\label{eqN0}
\end{equation}
and
\begin{equation}
N_{m\neq 0}=
\left (
\begin{array}{cc}
{{\lambda_b-\lambda_a}\over \lambda_a}a_m[(1+{{\lambda_b-\lambda_a}\over\lambda_a}
{\zeta_a\over \zeta})
G(p,z-z{'})+G({\hat p}_m,z-z{'})]\big |_{z=z{'}=0}+\phi_m&\;~~~{{\lambda_b-\lambda_a}\over \lambda_a}
a_m
G(p,z-z{'})\big |_{z=d,z{'}=0}\\[3mm]
{{\lambda_b-\lambda_a}\over \lambda_a}a_m G({\hat p}_m,z{'}-z)\big |_{z{'}=d,z=0} &~~0
\end{array}
\right )
\end{equation}
\end{widetext}
\!\!\!with $p=\sqrt{p_x^2+p_y^2}$, ${\hat p}_m=\sqrt{(p_x+2\pi m/\zeta)^2+p_y^2}$, the two-point 
correlation function in lateral Fourier space 
\begin{equation}
G(p,z-z{'})={k_BT\over 2Kp}\,e^{-p|z-z{'}|},
\end{equation}
and 
\begin{equation}
\displaystyle\phi_m=\left({\lambda_b-\lambda_a\over \lambda_a}\right)^2
{\sum_{k=-\infty}^{\infty}{}^{\!\!\!\displaystyle '}}a_k a_{m-k}G({\hat p}_k,0)
\end{equation}
where the prime at the summation sign indicates that in the sum the term $k=0$ is excluded. 
The patterning function $a(x)$ is represented as a Fourier series 
$a(x)=\sum_{k=-\infty}^{\infty}a_ke^{2\pi ikx/\zeta}$ with
\begin{equation}
a_k={1\over L}\int_{-L/2}^{L/2} {\rm d}x\, a(x) \exp(-2\pi ikx/\zeta)=
{1\over \pi k}\sin (\pi k\zeta_a/\zeta),
\end{equation}
where $L$ is the extension of the system along the $x-$direction (see Fig.\,\ref{geo55.eps}). 
We mention that the patterning function $a(x)$ is coordinate dependent and the phase of the 
coefficients $a_k$ depends on the choice of the position of the coordinate origin used for
a(x), but as expected the final result for the force is independent of this choice. We have 
checked this numerically [c.f., Eq.\,(\ref{eqF})].

Each pair of the momenta $({\bf p},{\bf q})$ indicates one element of the matrix $M$ which is
a $(2\times 2)$ matrix itself [Eq.\,(\ref{eq2})]. Although $M$ has an infinite number of elements
and is not diagonal it can be brought into a block diagonal form by an even number of row and 
column permutations~\cite{emig,emig1}. We take the system to be periodically extended 
with period $L=N\zeta$ along the $x$-axis with $N$ being an integer.
This leads to momenta $p_x$ that are integer multiples of $2\pi/L$. 
For $p_x$ fixed the factor $\mbox{\large $\delta$}\big(p_x+q_x+2\pi\,m/\zeta\big)$ in 
Eq.\,(\ref{eq2}) leads to nonvanishing matrix elements for all $q_x=-p_x-2\pi m/\zeta$
with $m\in \mathbb{Z}$. This allows one to identify the block structure of the matrix $M$.
For fixed $j$ the momenta $p_x=2\pi j/L+2\pi l/\zeta$ and $q_x=-2\pi j/L-2\pi k/\zeta$ 
form the block $M_j$ where $l, k\in \mathbb{Z}$ and $j=1, \cdots, N=L/\zeta$ so that there is 
no multiple counting of the momenta.
One can read off the elements of the infinite-dimensional 
block matrices $M_{j}(p_y,q_y)$ from Eq.\,(\ref{eq2}): 
\begin{equation}
M_{j,kl}(p_y,q_y)=2\pi\,\mbox{\large $\delta$}\big(p_y+q_y\big)\,B_{kl}\left({2\pi\,j\over L},
p_y\right)
\end{equation}
with the matrix $B$ given by
\begin{widetext}

\begin{equation}
B\left(p_x={2\pi\over L}\,j\right)=\left(
\begin{array}{ccccc}
\ddots&&\vdots&&\reflectbox{$\ddots$}\\[-0.3mm]
&N_0(p_x-{2\pi\over\zeta})&N_{-1}(p_x)&N_{-2}(p_x+{2\pi\over\zeta})&\\
\cdots&N_1(p_x-{2\pi\over\zeta})&N_0(p_x)&N_{-1}(p_x+{2\pi\over\zeta})&\cdots\\
&N_2(p_x-{2\pi\over\zeta})&N_1(p_x)&N_0(p_x+{2\pi\over\zeta})&\\[-1mm]
\reflectbox{$\ddots$}&&\vdots&&\ddots
\end{array}
\right),
\end{equation}
\end{widetext}
so that the matrix element $B_{kl}$ reads
\begin{equation}
B_{kl}\left({2\pi\,j\over L},p_y\right)=N_{m=k-l}\left({2\pi\,j\over L}+
{2\pi\,l\over\zeta},p_y\right).
\label{eqB}
\end{equation}
Note that a reindexation of the $(p_y,q_y)$-subspace is not necessary as $M$ is diagonal with 
respect to it [Eq.\,(\ref{eq2})].

Now the value of the path integral [Eq.\,(\ref{path})] given by
\begin{equation}
Z={\cal N}(det \, M)^{-1/2}
\label{ZdetM}
\end{equation}
can be calculated in Fourier space. The free energy $-k_BT\ln Z$ of the system usually lends 
itself to a decomposition into bulk, surface, and finite-size contributions~\cite{aside3}. 
The term $-k_BT\ln {\cal N}$ leads to the bulk free  energy. The factor ${\cal N}$ introduced
in Eq.\,(\ref{path}) is given by ${\cal N}=[det \, G^{-1}({\bf r},{\bf r{'}})]^{-1/2}$ where 
the determinant of the inverse of the two-point correlation function $G({\bf r}-{\bf r}{'})$ 
in an unperturbed nematic is calculated in the space actually occupied by the nematic, i.e., the
volume $V$. Thus the result for the bulk free energy is given by 
$F_{\rm bulk}=k_BTV\int_{0}^{Q_{max}} {{\rm d}Q\over (2\pi)^2}Q^2\ln \left({KQ_{\rm max}^{-3}Q^2\over k_BT}\right)$
where $Q_{\rm max}$ is an ultraviolet momentum cutoff of the order of the inverse size of the 
nematic molecules. The remaining part of the free energy 
$F=A(F_{\rm surf}^{z=0}+F_{\rm surf}^{z=d})+\delta F(d)$ 
is given by $F={k_BT \over 2}\ln\det M$. Here the $d$-independent terms $F_{\rm surf}^{z=0}$ 
and $F_{\rm surf}^{z=d}$ are the surface tensions associated with the interfaces at $z=0$ and
$z=d$, respectively, and the finite-size contribution $\delta F(d)$ is the fluctuation-induced
interaction. In the present model there are no other finite-size contributions.
Using the block structure of $M$, we obtain
$\ln\det M=\ln(\prod _{j=1}^{N}{\rm det}\,M_{j})=
\sum_{j=1}^{N}\ln({\rm det}\, M_j)\;$.
Thus the fluctuation-induced force ${\cal F}=-\partial_d F$ equals 
${\cal F}=-{k_BT \over 2}\sum_{j=1}^{N}{\rm Tr}\,(M_j^{-1}\partial_d M_j)$. 
The final result for the force reads
\begin{eqnarray}
{\cal F}=-{k_BTA\over 2\pi^2}&\mbox{\Large $\int$}_{0}^{\infty}&\!\!{\rm d}p_y\int_{0}^{2\pi/\zeta}\!\!{\rm d}p_x\nonumber\\
\times\!\!\!&{\rm Tr}&\!\!\!
\big(B^{-1}(p_x,p_y)\partial_d B(p_x,p_y)\big)\;,
\label{eqF}
\end{eqnarray}
where we have carried out the thermodynamic limit $L\rightarrow \infty$ so that the summation 
over $j$ is replaced by ${L\over 2\pi}\int_{0}^{2\pi/\zeta}{\rm d}p_x$. 
The trace over the continuous momenta $p_y$ is also replaced by 
${L\over \pi}\int_{0}^{\infty}{\rm d}p_y$ and $A=L^2$.
Equation\,({\ref{eqF}}) takes also into account that there are two independent fluctuating
fields $\nu$ (i.e., $\delta_x$ and $\delta_y$) which lead to a doubling of the force. Note that
in Eq.\,(\ref{eqF}) the trace is taken with respect to the remaining discrete 
indices $k$ and $l$ [Eq.\;(\ref{eqB})].
This can be calculated numerically by truncating the matrix
$B_{kl}$ at order $I$, i.e., $k,l=-(I-1)/2, \dots, 0,\dots, (I-1)/2$.  
The force ${\cal F}$ follows from extrapolating $I\rightarrow \infty$. 
\section{results}
In the limit $d/\zeta\gg 1$ the contributions from the matrices $N_m$ to ${\rm det}\,M$ decrease 
rapidly with increasing $m$ [Eq.~(\ref{eq2})]. In this limiting case it is sufficient to consider
only the contribution from $N_0(p_x,p_y)$, in the sense that truncating the 
matrix $B$ [Eq.\,(\ref{eqB})] at $I>1$ leaves the integrand in Eq.\,(\ref{eqF}) 
practically unchanged. Therefore it is instructive first to focus on this limiting case, which may
correspond to a nano-patterned substrate facing a homogeneous substrate at a micrometer 
separation, and to investigate analytically the behavior of the force. 
\subsection{Pattern of different anchoring strengths}
\label{II}
Here we consider a pattern characterized by stripes of homeotropic anchoring so that both $W_a$ and
$W_b$ are positive. In this case the fluctuations are suppressed at the substrates [Eq.~(\ref{eq3})]. Using Eqs.\,(\ref{eqB}), \,(\ref{eqN0}), and (\ref{eqF}) we find for the force
\begin{eqnarray}
{\cal F}(\!\!&d&\!\!\gg \zeta)=\lim_{\zeta\to0}{-k_BTA\over 2\pi^2}\int_{0}^{\infty}{\rm d}p_y\int_{0}^{2\pi/\zeta}{\rm d}p_x\nonumber\\&\times&\!\!{\rm Tr}\,\big(N_0^{-1}(p_x,p_y)\partial_d N_0(p_x,p_y)\big)\nonumber\\[3mm]
&=&{k_BTA\over \pi d^3}\int_{0}^{\infty}\!\!{\rm d}x\,{x^2\over
{x+d/\lambda_{\rm eff}\over x-d/\lambda_{\rm eff}}\exp {(2x)}+1}
\label{eq7}
\end{eqnarray}
where
\begin{equation} 
\lambda_{\rm eff}={\zeta \lambda_a \lambda_b\over \zeta_a\lambda_b+\zeta_b\lambda_a}
\label{eq7e}
\end{equation}
is introduced as an effective extrapolation length and $x=pd$ is the rescaled momentum. Thus in the
limit $d\gg \zeta$ the patterned substrate can be described by an effective anchoring energy per 
area with the force found between two homogeneous substrates where one substrate is characterized
by strong anchoring and the other substrate is characterized by a finite anchoring, i.e., a finite
extrapolation length $\lambda_{\rm eff}$ \cite{primozz}.
\begin{center}
\begin{figure}
\includegraphics[height=2.92\linewidth,angle=0]
{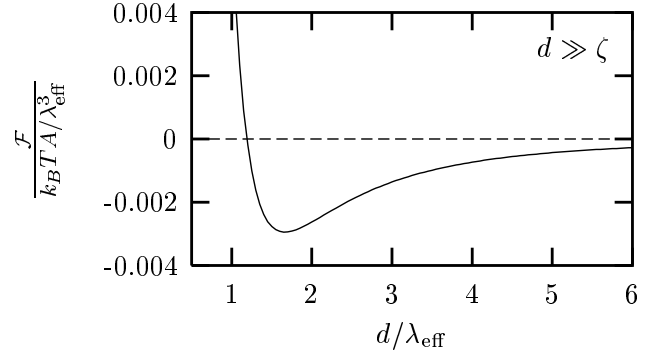}
\vspace{-20.0cm}
\caption{The fluctuation-induced force ${\cal F}$ as a function of the reduced separation $d/\lambda_{\rm eff}$ in
the case in which the patterned substrate can be described by an effective homeotropic
anchoring ($d\gg \zeta$, see Eq.\,(\ref{eq7})).  
A crossover from repulsion to attraction occurs for a 
separation $d/\lambda_{\rm eff}\simeq 1.19$ followed by a minimum at 
$({d\over \lambda_{\rm eff}},{{\cal F}\over k_BTA/\lambda_{\rm eff}^3}
)\simeq(1.65,-0.003)$.}
\label{geo5.eps}
\end{figure}
\end{center}

If $\lambda_a$ or $\lambda_b$ is zero, $\lambda_{\rm eff}$ vanishes and the force is 
long-ranged and attractive: 
\begin{equation}
{\cal F}(d\gg \zeta, \lambda_{\rm eff}=0)=-{k_BTA\,\zeta(3)\over 4\pi d^3}
\label{long}
\end{equation}
where $\zeta(s)=\sum_{t=1}^{\infty}t^{-s}$ is the Riemann zeta function. This expression equals 
the one obtained for substrates both characterized by homogeneous strong anchoring~\cite{ajdari}.
This implies that in this limit the fluctuation-induced force is not affected 
by the stripes with weaker anchoring.
This behavior resembles the one of the electrodynamic Casimir force between a flat and a 
rectangularly corrugated substrate~\cite{emig} in the limiting case that the periodicity is much
smaller than the amplitude $2h$ of the corrugation and the mean separation $H$ between the 
substrates. In this case the force equals the one between two flat substrates at a reduced mean 
separation $H-h$ so that the force is not affected by the valleys of the corrugated substrate.

\begin{center}
\begin{figure}
\includegraphics[height=2.92\linewidth,angle=0]
{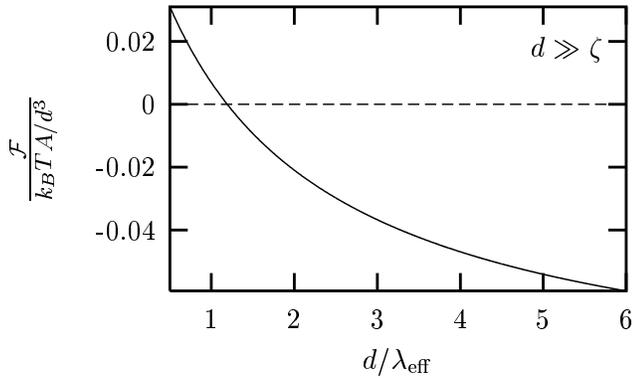}
\vspace{-20.0cm}
\caption{Amplitude of the fluctuation-induced force ${\cal F}$ relative to asymptotic 
behavior $d^{-3}$ as a function of the reduced separation $d/\lambda_{\rm eff}$ in
the case in which the patterned substrate can be described by an effective homeotropic
anchoring ($d\gg \zeta$, see Eq.\,(\ref{eq7})).
At $d/\lambda_{\rm eff}=0$ this amplitude attains the value $3\zeta(3)/(16\pi)\simeq 0.072$ 
linearly [Eq.\,({\ref{eq*}})] and approaches the value $-\zeta(3)/(4\pi)\simeq
-0.096$ for $d/\lambda_{\rm eff}\rightarrow \infty$ [Eq.\,(\ref{eqAt})].}
\label{geoA.eps}
\end{figure}
\end{center}

\begin{center}
\begin{figure}
\includegraphics[height=2.7\linewidth,angle=0]
{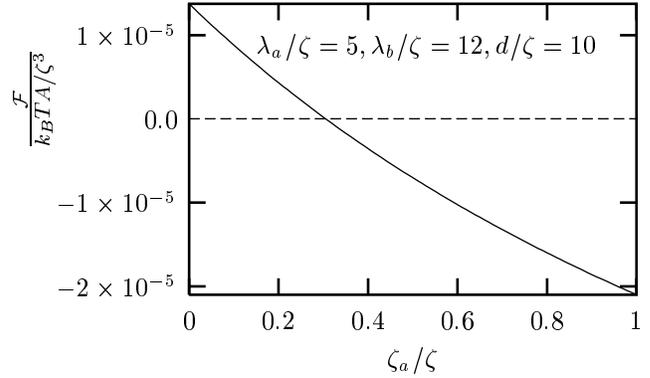}
\vspace{-17.70cm}
\caption{The fluctuation-induced force ${\cal F}$ as a function of the patterning 
ratio $\zeta_a/\zeta$ for a fixed reduced separation $d/\zeta$ (fulfilling the requirement
$d\gg \zeta$, see Fig.~\ref{geo5.eps}) and fixed reduced extrapolation lengths
$\lambda_{a(b)}/\zeta$ [Eq.\,(\ref{eq7})]. The crossover from repulsion to attraction indicates 
that upon increasing $\zeta_a/\zeta$ the system transforms from the effective strong-weak to the 
effective strong-strong anchoring regime.
For the given set of the parameters ($\lambda_a/\zeta=5$, $\lambda_b/\zeta=12$, and $d/\zeta=10$), 
the force becomes attractive when more than $31\%$ of the substrate consists of 
strong anchoring parts with $W_a>W_b$, i.e., $\lambda_a=K/W_a<\lambda_b=K/W_b$. Note that here 
${\cal F}$ is measured in units of $\zeta^3$ instead of $\lambda_{\rm eff}^3$ as 
in Fig.\;\ref{geo5.eps}.}
\label{geo6.eps}
\end{figure}
\end{center}

At intermediate values of $\lambda_{\rm eff}$ there is a crossover from attraction to 
repulsion (Fig.~\ref{geo5.eps}). At separations smaller than 
$\lambda^*\simeq 1.19~\lambda_{\rm eff}$, the boundaries effectively act as being homogeneous but
dissimilar\,\,--\,\,one boundary characterized by strong anchoring and the other boundary 
characterized by a finite weak anchoring so that the force is repulsive~\cite{ziherl}.
The asymptotic behavior of the force for $d/\lambda_{\rm eff}\ll 1$ is given by
\begin{equation} 
{\cal F}(\zeta\ll d\ll\lambda_{\rm eff})\approx {3 k_BTA\zeta(3)\over 16\pi d^3}\left(1-
{8\ln 2\over 3\zeta(3)}{d\over \lambda_{\rm eff}}\right).
\label{eq*}
\end{equation}
Thus in this regime the leading long-ranged repulsion term ${\sim}\,d^{-3}$, corresponding to two 
homogeneous substrates characterized by infinitely strong and zero anchoring (Dirichlet-Neumann 
boundary conditions), is weakened. At separations larger than $\lambda^*$, the boundaries 
effectively act as being similar\,\,--\,\,one boundary characterized by infinitely strong and 
the other boundary characterized by finite yet strong anchoring; therefore the force is 
attractive~\cite{ziherl}. The asymptotic behavior of the force for $d/\lambda_{\rm eff}\gg 1$ is 
given by (compare Eq.\,(\ref{long}))
\begin{equation}
{\cal F}(d\gg\lambda_{\rm eff},\zeta)\approx -{k_BTA\zeta(3)\over 4\pi d^3}
\left(1-{3\lambda_{\rm eff}\over d}\right).
\label{eqAt}
\end{equation}
This means that the long-ranged attraction ${\sim}\,d^{-3}$, corresponding to two homogeneous
substrates characterized by infinitely strong anchoring, is reduced. In Fig.\,\ref{geoA.eps} the
amplitude of the fluctuation-induced force ${\cal F}(d\gg\zeta)$, as given by the full expression
in Eq.\,(\ref{eq7}), is shown as a function of the reduced separation $d/\lambda_{\rm eff}$. 
In Fig.\,\ref{geo6.eps} the force [Eq.\,(\ref{eq7}] is shown as a function of the patterning 
ratio $\zeta_a/\zeta$. Also in this case the above considerations provide an understanding of 
the crossover from repulsion to attraction upon changing the patterned substrate from the
effectively weak to the effectively strong anchoring regime.

In the opposite limiting case $d/\zeta\ll 1$, the force is given by
\begin{eqnarray}
{\cal F}(d\ll \zeta)={k_BTA\over \pi d^3}\Bigg({\zeta_a\over \zeta}\int_{0}^{\infty}{\rm d}x
{x^2\over{x+d/\lambda_{a}\over x-d/\lambda_{a}}\exp {(2x)}+1}\nonumber\\
+{\zeta_b\over \zeta}\int_{0}^{\infty}
{\rm d}x{x^2\over
{x+d/\lambda_{b}\over x-d/\lambda_{b}}\exp {(2x)}+1}\Bigg).\
\label{eqA}
\end{eqnarray}
Here the geometrically weighted average is analogous to the results obtained within the proximity
force approximation \cite{pfa}. This scheme amounts to using the force density obtained from 
the homogeneous case and integrating over the local contributions of the force densities 
corresponding to the regions with anchoring energies per area $W_a$ and $W_b$, respectively.
The particular contributions from the regions close to the chemical steps cause deviations
from this approximation. Therefore this result holds only for a low number density of chemical 
steps, i.e., for $d\ll \zeta$.

\begin{figure}
\includegraphics[height=2.92\linewidth,angle=0]
{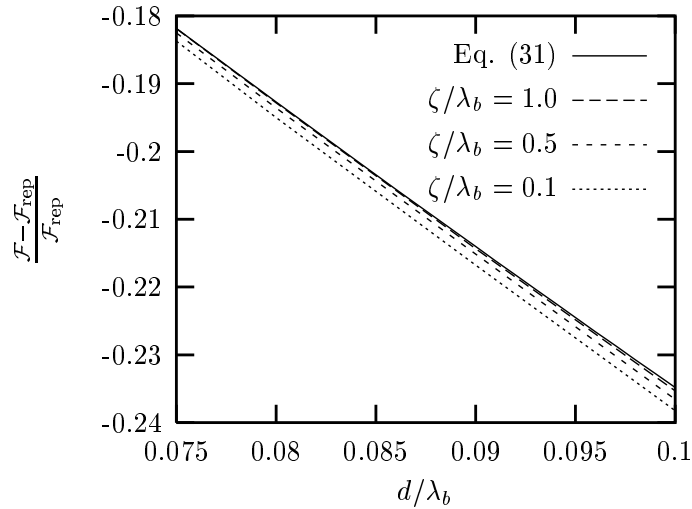}
\vspace{-18.0cm}
\caption{Rescaled relative decrease of the fluctuation-induced force in the case in which the 
pattern is characterized by alternating anchoring strengths as a function of the reduced
separation  $d/\lambda_b$ compared to the repulsive force 
${\cal F}_{\rm rep}=3 k_BTA\zeta (3)/(16 \pi d^3)$ between two homogeneous boundaries characterized
by infinitely strong and zero anchoring energy, respectively. For $d\ll \zeta$ Eq.\,(\ref{eqA}) 
holds. Both integrals in Eq.\,(\ref{eqA}) happen to have the same form as in Eq.\,(\ref{eq7}) 
(which holds, however, for $d\gg \zeta$) with $\lambda_{\rm eff}$ 
replaced by $\lambda_a$ and $\lambda_b$, respectively. According to the discussion in the second 
paragraph following Eq.\,(\ref{eq7}), for the parameter values 
$\zeta_a=\zeta_b$ and $\lambda_a/\lambda_b=0.4$ chosen here these integrals are positive for
$d/\lambda_b<0.48$ for the contribution associated with $\lambda_a$ and $d/\lambda_b<1.19$ for the
contribution associated with $\lambda_b$. Thus one expects that on the basis of Eq.\,(\ref{eqA})
${\cal F}$ reduces to ${\cal F}_{\rm rep}$ as given by Eq.\,(\ref{eq*}) for $d/\zeta\rightarrow 0$. 
The solid curve corresponds to ${\cal F}$ given by Eq.\,(\ref{eqA}) and the other curves correspond
to the full numerical results obtained from Eq.\,(\ref{eqF}) for the indicated values 
of $\zeta/\lambda_b$. All curves appear to vanish for $d/\zeta\rightarrow 0$ and thus confirm the
above expectation. Moreover, by increasing the periodicity the difference between the solid curve
and the numerical results vanishes and thus confirms Eq.\,(\ref{eqA}) as a reliable approximation in
the limit $d/\zeta\rightarrow 0$.}
\label{Q}
\end{figure}

\begin{figure}
\includegraphics[height=2.92\linewidth,angle=0]
{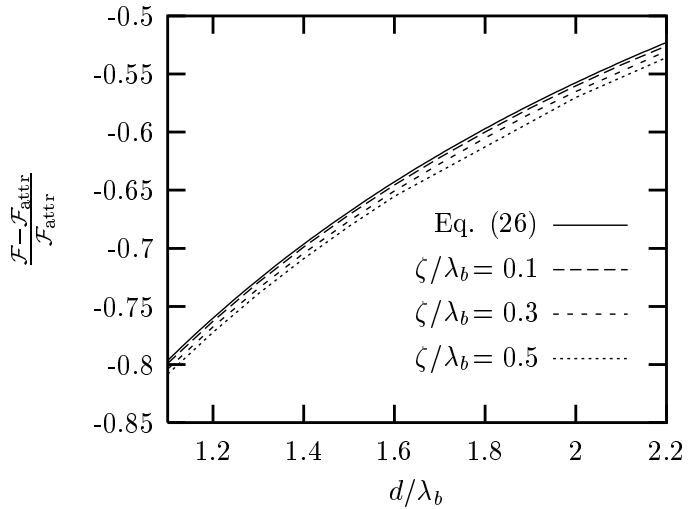}
\vspace{-18.cm}
\caption{Rescaled relative decrease of the fluctuation-induced force in the case in which the
pattern is characterized by alternating anchoring strengths as a function of the reduced
separation  $d/\lambda_b$ compared to the attractive force 
${\cal F}_{\rm attr}=-k_BTA\zeta (3)/(4 \pi d^3)$ between two homogeneous substrates characterized
by infinitely strong anchoring energies. For $d\gg \zeta$ Eq.\,(\ref{eq7}) holds.
According to the discussion in the second paragraph following Eq.\,(\ref{eq7}) this integral
is negative for $d/\lambda_b>0.68$ for the parameter values $\zeta_a=\zeta_b$ 
and $\lambda_a/\lambda_b=0.4$ chosen here. Thus one expects that ${\cal F}$ reduces to
 ${\cal F}_{\rm attr}$ given by Eq.\,(\ref{eqAt}) for $\zeta/d\rightarrow 0$. 
The solid curve corresponds to ${\cal F}$ given by Eq.\,(\ref{eq7}) and the other curves 
correspond to the full numerical results obtained from Eq.\,(\ref{eqF}) for the indicated values
of $\zeta/\lambda_b$. All curves appear to vanish for $\zeta/d\rightarrow 0$ and thus confirm
the above expectation. Moreover, by decreasing the periodicity $\zeta$ the difference between
the solid curve and the full numerical results vanishes and thus confirms 
Eq.\,(\ref{eq7}) as a reliable approximation in the limit $\zeta/d\rightarrow 0$.}
\label{Q1}
\end{figure}

\vspace{-0.8cm}
For intermediate values of $d/\zeta$ we have calculated the force ${\cal F}$ numerically based on 
the complete expression given by Eq.\,(\ref{eqF}). In order to highlight the deviation of the 
fluctuation-induced force in the considered patterned system from the long-ranged behavior in the
homogeneous cases, we show in Figs.\,\ref{Q} and \ref{Q1} the rescaled decrease of the force 
relative to the decay ${\sim}\,d^{-3}$ in the repulsion and attraction regime, respectively.
\subsection{
Pattern of competing anchoring energies}
\label{III}
In the following we generalize the system to a patterned bottom substrate characterized not only by
different anchoring strengths but also different preferred molecular axes.
We consider a pattern consisting of alternating stripes of either homeotropic or degenerate planar
anchoring, i.e., positive values for $W_a$ and negative values for $W_b$, so that the liquid crystal
is subject to competing preferred orientations at the bottom substrate while the upper substrate 
still exhibits strong homeotropic anchoring. In this case fluctuations are suppressed at the stripes
of homeotropic anchoring and enhanced at the stripes of planar anchoring [Eq.~(\ref{eq3})].
For those ranges of the model parameters for which the frustrating effect of the interlaced planar
anchoring does not modify the mean orientation of the director, the uniform configuration of the 
director is thermodynamically stable and the system can be described as in the previous subsection.
However, if the planar anchoring becomes dominant, the uniform director configuration is destabilized.
For this case, we determine the behavior of 
the fluctuation-induced force before the onset of the ensuing structural phase transition. 
\begin{center}
\begin{figure}
\includegraphics[height=2.92\linewidth,angle=0]
{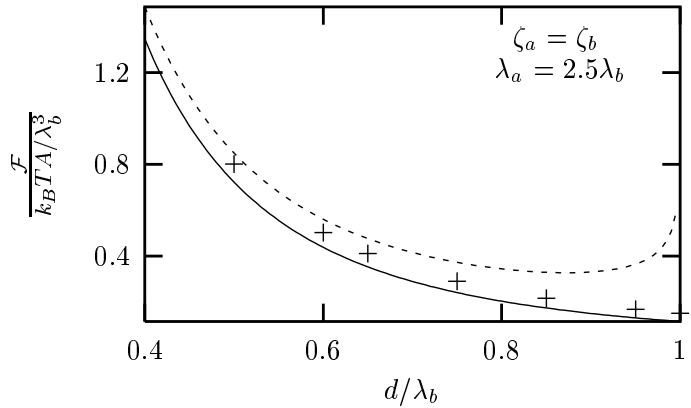}
\vspace{-19.cm}
\caption{The fluctuation-induced force as a function of the reduced separation $d/\lambda_{b}$ in
the case in which the pattern is characterized by competing anchoring energies. The solid curve
shows the result for small periodicity ($\zeta\ll d$) where the pattern can be described by an 
effective anchoring energy per area 
$W_{\rm eff}=\zeta_a W_a/\zeta+\zeta_b W_b/\zeta <0$ [Eq.\,(\ref{hybrid1})]. The dashed curve shows
the result for large periodicity ($\zeta\gg d$) for which the proximity force approximation is
valid~[Eq.\,(\ref{eqp})]. The data points (+) show our full numerical results [Eq.\,(\ref{eqF})] for
$\zeta/\lambda_{b}=1.5$. This illustrates that as expected ${\cal F}$ approaches the proximity force 
solution (dashed curve) for
$d/\zeta\rightarrow 0$ and the effective anchoring solution (solid curve) for
$d/\zeta\rightarrow \infty$.}
\label{pata0.ps}
\end{figure}
\end{center}

\vspace{-0.8cm}
We consider again first the case $d/\zeta\gg 1$ which allows us to describe the force in terms of 
an effective anchoring energy and an effective extrapolation length. Within the same approximation
as in the preceding subsection, i.e., truncating the matrix $B$ at order $I=1$, the force
[Eq.\,(\ref{eqF})] is given by
\begin{equation}
{\cal F}(d\gg \zeta)={k_BTA\over \pi d^3}\int_{0}^{\infty}{\rm d}x{x^2\over
\,{x-d/\lambda_{\rm eff}\over x+d/\lambda_{\rm eff}}\exp {(2 x)}+1}
\label{hybrid1}
\end{equation}
where 
\begin{equation}
\lambda_{\rm eff}=\zeta \lambda_a \lambda_b/\mid \zeta_a\lambda_b-\zeta_b\lambda_a\mid
\label{hybrid1e}
\end{equation}
and $x=p d$ is the rescaled momentum~\cite{aside2}. The form of this force is that of the force
for the hybrid cell~\cite{karimi} with the bottom substrate described by an effective planar 
anchoring $W_{\rm eff}=\zeta_a W_a/\zeta+\zeta_b W_b/\zeta<0$. Thus in this limit the patterning
enters only via this expression for $W_{\rm eff}$.

For $\zeta\ll d \ll \lambda_{\rm eff}$ the long-ranged repulsion between the boundaries,
characterized by strong and effectively weak anchoring, is enhanced:
\begin{equation}
{\cal F}(\zeta\ll d \ll \lambda_{\rm eff})\approx {3k_BTA\zeta (3)\over 16\pi d^3}\left(1+
{8\ln 2\over 3\zeta (3)}{d\over \lambda_{\rm eff}}\right).
\end{equation}
Upon approaching the critical separation $d_c=\lambda_{\rm eff}$ the force increases and, within
the Gaussian approximation, diverges logarithmically at $d_c$:
\begin{equation}
{\cal F}(d/d_c\rightarrow 1)\approx -
{3k_BTA\over 4\pi \lambda^3_{\rm eff}}\left[ \ln \Big(1-{d\over d_c}\Big) + \ln 2 +...\right]\;.
\end{equation}
This behavior follows from the fact that for $d=\lambda_{\rm eff}$ the denominator of the integrand
in Eq.\,(\ref{hybrid1}) vanishes as ${2 \over 3} x^3$ for $x\to0$. Thus the pretransitional 
behavior $(d\rightarrow d_c)$ of the force is related to the sigularity of the soft mode
$(x\rightarrow 0)$.

The logarithmic divergence is a characteristic feature of the perturbative 
method [Eq.\,(\ref{eq1p})] if applied to a system subject to structural phase 
transitions ~\cite{karimi,karimi1,nariya,bostjan}. Here the planar anchoring destabilizes 
the uniform structure governed by the strong homeotropic anchoring at the upper boundary. Upon
increasing the separation $d$ the influence of the upper boundary decreases and the destabilizing
effect of the substrate characterized by the planar anchoring increases so that upon approaching
$d=d_c$ the uniform structure becomes unstable~\cite{barberi}. Beyond the critical separation, the
director structure is no longer uniform and the Gaussian Hamiltonian given by Eq.\,(\ref {eq1}) does
no longer describe the system. For $d\gtrsim d_c$ one must consider the fluctuations around a 
nonuniform configuration ${\bf n}_0({\bf r})$. It is possible and even likely that the actual 
critical film thickness for the structural phase transition is not given by $d_c=\lambda_{\rm eff}$ 
as obtained in the present perturbative approach. 

In the regime of the uniform director configuration, for separations comparable to the periodicity
we have calculated the force numerically. As for Eq.\,(\ref{eqA}), in the limit $d/\zeta\ll 1$ the 
force is given by the geometrically weighted average of the local force densities:
\begin{eqnarray}
{\cal F}(d\ll \zeta)={k_BTA\over \pi d^3}\Bigg({\zeta_a\over \zeta}\int_{0}^{\infty}{\rm d}x
{x^2\over {x+d/\lambda_{a}\over x-d/\lambda_{a}}\exp {(2x)}+1}\nonumber\\
+{\zeta_b\over \zeta}\int_{0}^{\infty}
{\rm d}x{x^2\over
{x-d/\lambda_{b}\over x+d/\lambda_{b}}\exp {(2x)}+1}\Bigg).\
\label{eqp}
\end{eqnarray}
In Fig.~\ref{pata0.ps} the behavior of the fluctuation-induced force is shown for the three 
different regimes. 
\section{summary and conclusion}
\label{IV}
Based on the Frank [Eq.\,(\ref{eq0})] and Rapini-Papoular [Eq.\,(\ref{eq6})] expression for the
bulk and surface free energy, respectively, we have studied by field theoretical techniques the
contribution of Gaussian fluctuations of the orientational order to the effective force acting on
the planar and parallel boundaries of a nematic film of thickness $d$. The upper boundary exhibits
strong and homogeneous homeotropic anchoring whereas the bottom boundary is characterized by a 
one-dimensional, steplike periodic modulation with period $\zeta$ of either the strength or the
orientation of the anchoring which is homeotropic or planar (see Fig.~\ref{geo55.eps}). The system
parameters are chosen such that the thermal average of the nematic order configuration is 
spatially homogeneous throughout the film. Our main results are the following: 

(1) For $d\gg \zeta$, the fluctuation-induced force is proportional to $d^{-3}$ times a scaling 
function which attains the same form as in the case that the patterned bottom substrate exhibits a 
homogeneous anchoring energy per area $W_{\rm eff}=\zeta_a W_a/\zeta+\zeta_b W_b/\zeta$ with the 
different anchoring energies per area $W_{a(b)}=K/\lambda_{a(b)}$ weighted according to their 
lateral geometric contribution (Eqs.\,(\ref{eq7}) and (\ref{eq7e}) as well as (\ref{hybrid1}) and
(\ref{hybrid1e})). If a change of the pattern of the anchoring strength causes the corresponding
effective boundary condition on the patterned substrate to change from homeotropic weak to 
homeotropic strong anchoring, the force exhibits a crossover from repulsion to attraction 
(see Figs. \,\ref{geo5.eps},\,\ref{geoA.eps}, and \ref{geo6.eps}).
For a pattern of competing anchoring energies in the case that the stripes with planar anchoring
are dominant relative to the stripes of homeotropic anchoring, the fluctuation-induced force is 
repulsive and nonmonotonic. 

(2) For $d\ll \zeta$, the proximity approximation is valid, i.e., the force is given by the 
geometrically weighted average of the local force densities obtained from the homogeneous 
substrates (Eqs.\,(\ref{eqA}) and (\ref{eqp})).

(3) For intermediate separations, we have examined the behavior of the force numerically.
For a pattern of homeotropic anchoring with alternating strengths, the rescaled relative 
decrease of the fluctuation-induced force as a function of the reduced separation compared to 
the long-ranged force is summarized in Figs.\,\ref{Q} and \ref{Q1} for several values of the reduced
periodicity. We note that by increasing the reduced periodicity the system approaches the 
limit of the proximity force approximation given by Eq.\,(\ref{eqA}) 
(solid line in Fig.\,\ref{Q}), while by decreasing the reduced periodicity the system approaches 
the limit of the effective anchoring approximation given by 
Eq.\,(\ref{eq7}) (solid line in Fig.\,\ref{Q1}).

(4) For a pattern of competing anchoring energies, the behavior of the force is summarized in
Fig.\,\ref{pata0.ps}. Also in this case the full numerical solution interpolates between the 
asymptotic formulae given for small $d$ [Eq.\,(\ref{eqp})] and large $d$ [Eq.\,(\ref{hybrid1})].

Besides the force due to fluctuations of the director in a nematic film, there are additional forces
acting between interfaces of the film such as the well-known van der Waals dispersion 
force\,\cite{isra} so that the total force is the sum of the fluctuation-induced and the
dispersion forces. In the case that the mean director is inhomogeneous so-called structural 
forces appear in addition  which are not present in the homogeneous case considered here.  
For a film geometry of thickness $d$ and area $A$, the dispersion force decays as $-AH/(6\pi d^3)$ 
where $H$ is the so-called Hamaker constant. As implied by Eqs.\,(\ref{eq7}) and
(\ref{hybrid1}), the magnitude of the fluctuation-induced force scales as $k_BTA/d^3$. Thus, apart
from a numerical prefactor of order one, the overall ratio between the dispersion force and the 
fluctation-induced force is given by $H/k_{B}T$. Since liquid crystals and glass substrates 
have typically comparable indices of refraction, the Hamaker constant is of the 
order $10^{-21}J$\,\cite{kocevar} which is the same order of magnitude as $k_BT$ at room
temperature. Therefore typically the dispersion and fluctuation induced force in liquid crystals
are of the same order of magnitude and thus comparable. We mention that for a more adequate 
estimate of the Hamaker constant the anisotropy of the liquid crystal has to be taken into
account\,\cite{sarlah}. For larger thicknesses the dispersion force decays faster ($\sim 1/d^4$) due
to retardation\,\cite{ret}, so that for sufficiently thick films the fluctuation-induced force
dominates.

Finally we note that while the direct measurement of the fluctuation-induced force in liquid 
crystals has not yet been accomplished, these forces affect in a characteristic way the 
pattern formation of thin liquid-crystalline dewetting films\,\cite{valignat,her,zumer}.


\begin{thebibliography}{99}
\bibitem{aside1}
The cross terms in the expansion of Eq.\,(\ref{eq0}) in terms of $\delta_x$ and $\delta_y$ drop out 
after performing integrations by parts and assuming periodic boundary conditions. In the general case of an anisotropic elastic free energy, the Hamiltonian can be diagonalized in terms of longitudinal
and transverse components (see Ref.\,\cite{ajdari}):
$H[\phi_1,\phi_2]={1\over 2}\sum_{i=1}^{2}\int
{\rm d}^3x\,[K_i(\mbox{\boldmath $\nabla$}_{\perp} \phi_i)^2+K_3(\partial_z\phi_i)^2]$ where $\mbox{\boldmath $\nabla$}_{\perp}$ is the nabla operator with respect to the lateral coordinates.
Consequently, for this case the result for the force as obtained from Eq.\,(\ref{eq1}) has to be 
multiplied by ${1\over 2}\left({K_3\over K_1}+{K_3\over K_2}\right)$.
\bibitem{aside3} Note however, that there are cases such as the hybrid cell 
studied in Ref.\,\cite{karimi}
in which the surface and the finite-size 
free energies cannot be disentangled for long wavelength modes.
\bibitem{aside2} We exclude here the case of tilted easy directions for which the director structure
might not be uniform and, therefore, the formalism introduced here cannot be applied.
\end{thebibliography}
\end{document}